\documentclass[aps,prl,twocolumn,groupedaddress,superscriptaddress,showpacs,preprintnumbers,amsmath,amssymb]{revtex4-1}

\usepackage{graphicx}
\usepackage{dcolumn}
\usepackage{color}
\usepackage{bm}

\newcommand{\be}{\begin{equation}} \newcommand{\ee}{\end{equation}}
\newcommand{\bea}{\begin{eqnarray}} \newcommand{\eea}{\end{eqnarray}}

\begin{document}

\title{Outbreaks of coinfections: the critical role of cooperativity} 

\author{Li Chen}\affiliation{Max Planck Institute for the Physics of Complex Systems, Dresden, Germany}
\author{Fakhteh Ghanbarnejad}\affiliation{Max Planck Institute for the Physics of Complex Systems, Dresden, Germany}
\author{Weiran Cai}\affiliation{Faculty of Electrical and Computer Engineering, TU Dresden, Dresden, Germany}
\author{Peter Grassberger}\affiliation{Max Planck Institute for the Physics of Complex Systems, Dresden, Germany}
                          \affiliation{JSC, FZ J\"ulich, D-52425 J\"ulich, Germany}
 \email{chenli,fakhteh,grass@pks.mpg.de; weiran.cai@tu-dresden.de}

\date{\today}

\begin{abstract}
Modeling epidemic dynamics plays an important role in studying how
diseases spread, predicting their future course, and designing strategies to control them.
In this letter, we introduce a model of SIR (susceptible-infected-removed) type which
explicitly incorporates the effect of {\it cooperative coinfection}. More precisely, each individual
can get infected by two different diseases, and an individual already infected with one disease
has an increased probability to get infected by the other. Depending on the amount of this increase,
we prove different threshold scenarios. Apart from the standard continuous phase transition for
single-disease outbreaks, we observe continuous transitions where both diseases must coexist,
but also discontinuous transitions are observed, where a finite fraction of the population is
already affected by both diseases at the threshold. All our results are obtained in a mean field model
using rate equations, but we argue that they should hold also in more general frameworks.

\end{abstract}

\maketitle

\emph{Introduction.} --- From the Plague of Athens to the 14th century Black Death, the 1918-1919
Spanish flu, and to the recent HIV pandemic, infectious diseases  have caused more deaths than any other
factors, such as wars or famines \cite{Hays}. Mathematical models are thus extremely
important for understanding the outbreak and subsequent dynamics  of epidemics \cite{Anderson, Hethcote}.
Such models have been studied in particular by statistical physicists, who relied on the notion of
{\it universality} in critical phenomena to describe valid features of real epidemics in terms of
highly idealized and simplified models.

A pioneering work in this direction was carried out by Kermack and McKendrick \cite{KK}, who introduced in 1927 the
`Susceptible-Infective-Removed' (SIR) model, in which each individual can be in one of three states
(or ``compartments") S, I, and R. Infected individuals are ``removed" (i.e., recover or die) with fixed rate,
while susceptible ones can get infected with a rate that is proportional to the fraction of infecteds.
`Removed' individuals, finally, stay as they are and do not take part any more in the dynamics. When
treating this on a spatial grid with nearest-neighbor infection, starting with all sites being
susceptible except for one infected would lead to a percolation cluster of removed
sites \cite{Mollison,Grassberger}. As the infection rate passes through the percolation
threshold, the average relative cluster size increases gradually from zero, implying that the onset of
the epidemic is a continuous or ``second order" phase transition. In the mean field treatment of
\cite{KK}, basically the same is true: an infinitesimal fraction of initially
infected individuals will have no effect if the process is subcritical, while it leads to a finite fraction
of removed individuals if the threshold is passed. This fraction is zero at threshold and increases
continuously above it.

In recent years such models of epidemic spreading have been much studied on networks \cite{Newman,Dorogovtsev}.
Also, there was much interest in mechanisms that might lead to discontinuous phase transitions where
the epidemic involves a finite fraction of the epidemic already at threshold. Models that show (or were
claimed to show) the latter include ``explosive percolation" \cite{Achlioptas}, the Dodds-Watts model for
cooperative complex contagion \cite{Dodds} (see also \cite{Janssen,Bizhani,Goltsev}),
cascades on interdependent networks \cite{Buldyrev,Parshani,Woo}, models with long range infection 
\cite{Boettcher,grass-levy-1d}, and models with structured immunity \cite{Reluga}.

Surprisingly little work was, however, devoted in the statistical physics literature to the dynamics of 
multiple diseases. The competition between epidemics that are mutually exclusive or antagonistic was studied in
\cite{Newman-a,Funk10,Marceau11,Miller}.
But much more interesting is the case of {\it cooperative} multiple diseases, where the presence of one disease 
makes the other(s) more likely to spread. Such ``syndemics" \cite{Singer09} or ``coinfections" are well
documented in the epidemiological literature. Cases include the increased
incidence of tuberculosis during the 1918-1919 Spanish flu \cite{Brundage,Oei} and the fact that
persons infected by HIV have a higher risk to be infected by other pathogens, including
hepatitis B \& C \cite{Sulkowski}, TB \cite{Sharma} and Malaria \cite{Abu-Raddad}. 

In such cases, as in other cases of positive feed-back, one can 
expect much more violent outbreaks. Indeed, cooperative coinfections have been studied in the mathematics 
literature \cite{Abu-Raddad,Pilyugin,Marcheva}.
In \cite{Abu-Raddad} the case of HIV and malaria was modeled by a
compartmental model in terms of ODEs similar to Eq.(1), but the intention there was to
describe the syndemic as realistic as possible, introducing a large number of parameters and 
disregarding any phase transitions. Recently \cite{Newman13}, a model more in spirit of 
the present paper was proposed (albeit with completely different formalism). But it deals only with 
strongly asymmetric cases where only one of the diseases can
influence the other, while we are mostly interested in symmetrical cases with mutual cooperativity, 
where more interesting phenomena are expected. Closest in spirit to the present work are 
\cite{Pilyugin,Marcheva}. There it was shown, by using also ODEs similar to Eq.(1), that cooperativity
can lead to ``backward bifurcations", which are just first order mean field transitions in physics 
jargon. 

In the present letter, we propose what we believe to be the simplest SIR type model with two diseases 
(called $A$ and $B$) that leads to first order transitions. In this model, the infection rate for disease $A$ 
is increased, if the individual has or had disease $B$ and vice versa. When recovering from disease A, say, an
individual is `removed' from the population that is susceptible to A, but it still can be
infected by B. We shall only treat this model in mean field
approximation (described by rate equations similar to those in \cite{KK}).
Moreover, we shall mostly deal only with a very special case where there is symmetry between $A$
and $B$, and where present and past infections by $B$ have the same effect on infection by $A$.
In spite of these limitations we find a surprisingly rich behavior with two novel outbreak
mechanisms, one continuous and the other discontinuous.

\begin{figure}[tp]
\includegraphics[scale=0.39]{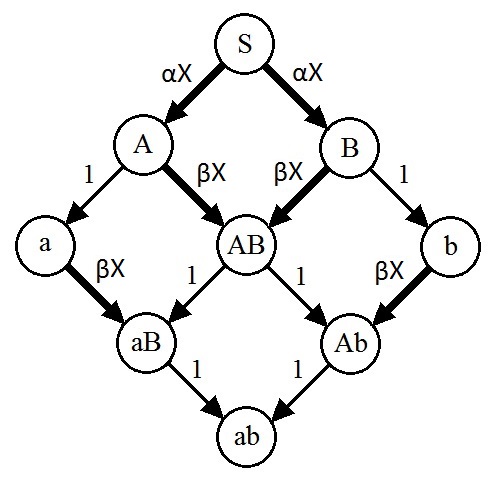}
\caption{Flow chart in two disease coinfection with $A,B$ symmetry and restrictions on the infection rates
   as discussed in the text. Capital letters $A$ and $B$ represent infective states,
   lower case letters $a$ and $b$ stand for `recovered' ones. Infecting neighbors are not
   indicated explicitely, but it is assumed that all individuals infected with disease $A$, say, have the
   same chances to pass $A$ on to another individual. Thus every infection process occurs with a rate
   proportional to the fraction $X$ of the population having the corresponding disease.}
\end{figure}

\emph{Model.} --- Consider a population of fixed size, where every individual can be in one of three
possible states -- susceptible, infective, and recovered/removed -- with respect to a each of
two diseases, called $A$ and $B$ in the following. This gives nine possible states for each
individual, denoted by $S,A,B,AB,a,b,aB,Ab$ and $ab$. Here capital letters refer to actual
infections, while lower-case letters refer to previous infections. Thus, e.g, a person in state
$aB$ has recovered from (and is thus immune to) disease $A$, but has presently disease $B$. Single
letters refer to states where the person is still susceptible with respect to the other disease.
We assume a well mixed population with normal first-order ``chemical" kinetics. Designing the
nine states by an index $i=0,\ldots 8$ and by $x_i$ the corresponding fraction (with $\sum_{i=0}^8x_i=1$),
the dynamics can thus be written as
\be
   \frac{dx_i}{dt} = \sum_j \mu_{ij}(x_j - x_i)  + \sum_{jk} \nu_{ijk} x_k (x_j - x_i),
\ee
where $\mu_{ij}$ is the rate with which state $i$ recovers spontaneously to state $j$ and $\nu_{ijk}$
is the rate for $i$ to change into $j$ due to infection by $k$.

In the following we shall make several simplifying assumptions:\\
(1) Diseases $A$ and $B$ have the same infection and recovery rates, and also the initial
    conditions are symmetric under the exchange $A\leftrightarrow B$.\\
(2) All infected states have the same recovery rate, which we set equal to one; state $AB$ cannot
    go directly to $ab$, but must first go to $aB$ or $Ab$.\\
(3) Infection rates for disease $A$, say, depend only on the fact whether the target has (or has had) $B$
    or not, but are independent of whether the infector has (had) $B$ or not. Thus we have only two
    different infection rates: Rate $\alpha$ for a target that is still susceptible for both diseases, and
    rate $\beta$ for targets which have or have had the other disease.

Thus we end up with the flow pattern depicted in Fig.~1. At the end of the paper we shall briefly
discuss more general cases where some of these restrictions are released.

Due to assumptions (1) and (3), all bilinear terms in Eq.~(1) are proportional to the fraction
\be
     X = [A] +[AB] + [Ab] = [B] + [AB] + [aB]
\ee
in the population that has the corresponding disease. Defining in addition
\be
    S = [S] \quad {\rm and} \quad P = [A] + [a] = [B] + [b],
\ee
Eq.~(1) can be rewritten as
\bea
    \dot{S} & = & -2\alpha SX \nonumber \\
    \dot{P} & = &(\alpha S -\beta P)X \nonumber \\
    \dot{X} & = &(\alpha S +\beta P)X -X.           \label{ODE3}
\eea

Thus we have been able to reduce our model to three ODEs with two control parameters $\alpha, \beta$.
The {\it cooperativity} is defined as the ratio $C=\beta/\alpha$. In particular, we are interested in the
$t\to\infty$ limit of solutions of Eq.~(\ref{ODE3}) with initial conditions
$S_0 = 1-\epsilon$ and $X_0 = P_0 = \epsilon/2$.
This corresponds to an initial population where most of the individuals (except for a small fraction
$\epsilon$) are susceptible to both diseases, while the rest has either $A$ or $B$. Including in the
initial state also recovered individuals or individuals with both diseases would not give more
insight. For $t\to\infty$ all activity has to die out, whence $X_\infty = 0$. Our ``order parameter"
is the asymptotic fraction $R=1-S_\infty$ of the population that has had at least one of the two diseases.
We expect interesting phenomena when $C>1$, since only then $X$ can have an intermediate growth phase
even when the single-disease infection rate $\alpha$ is smaller than 1. For $C=1$ the two diseases evolve
independently, and for $0<C<1$ we expect only minor modifications of the threshold behavior from independence.

\emph{Numerical Results.} --- In Fig. \ref{fig.2}, we show results obtained by integrating Eqs.~(\ref{ODE3})
numerically. We see the following main features:

\begin{figure}[htp]
\includegraphics[scale=0.22]{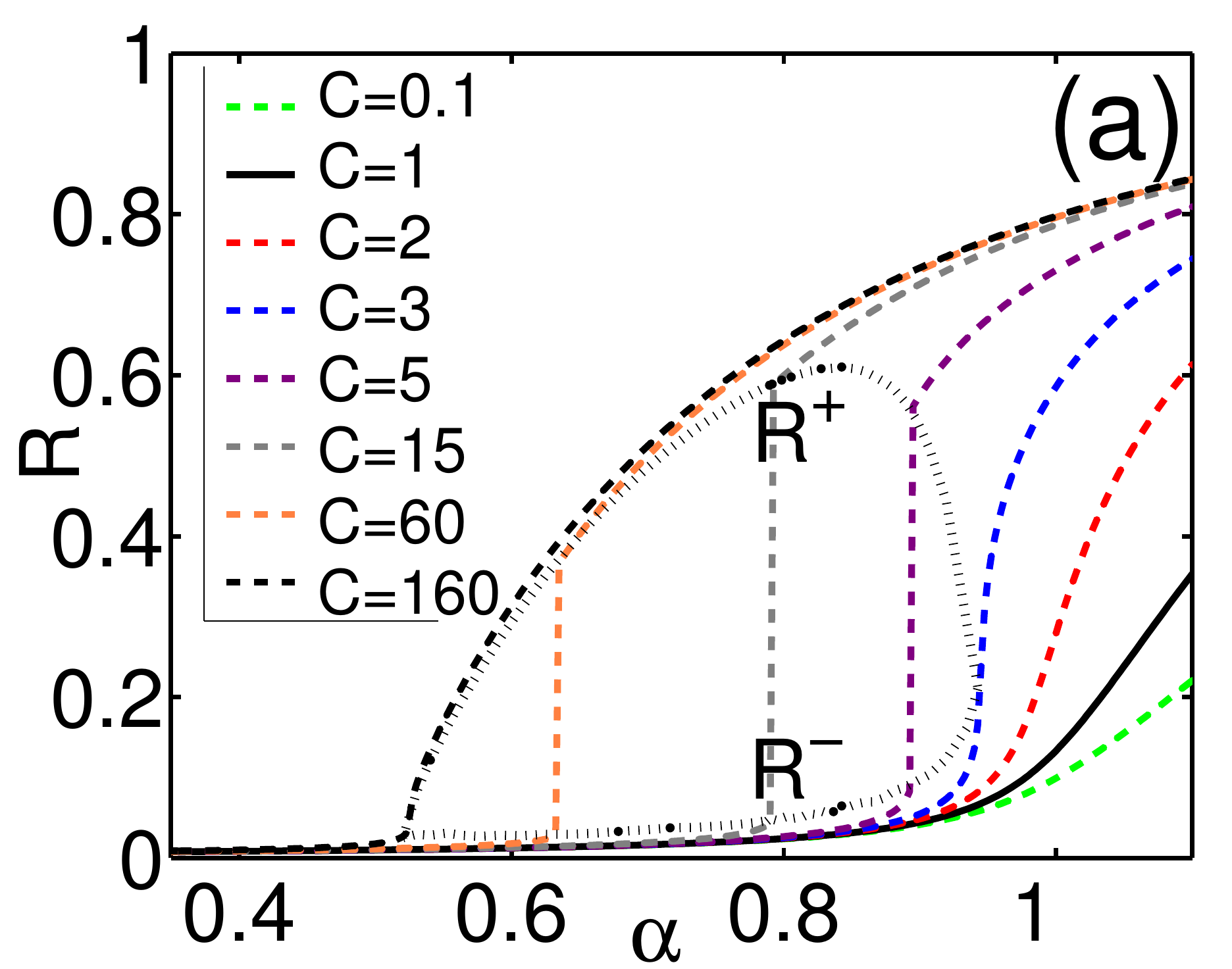}
\includegraphics[scale=0.22]{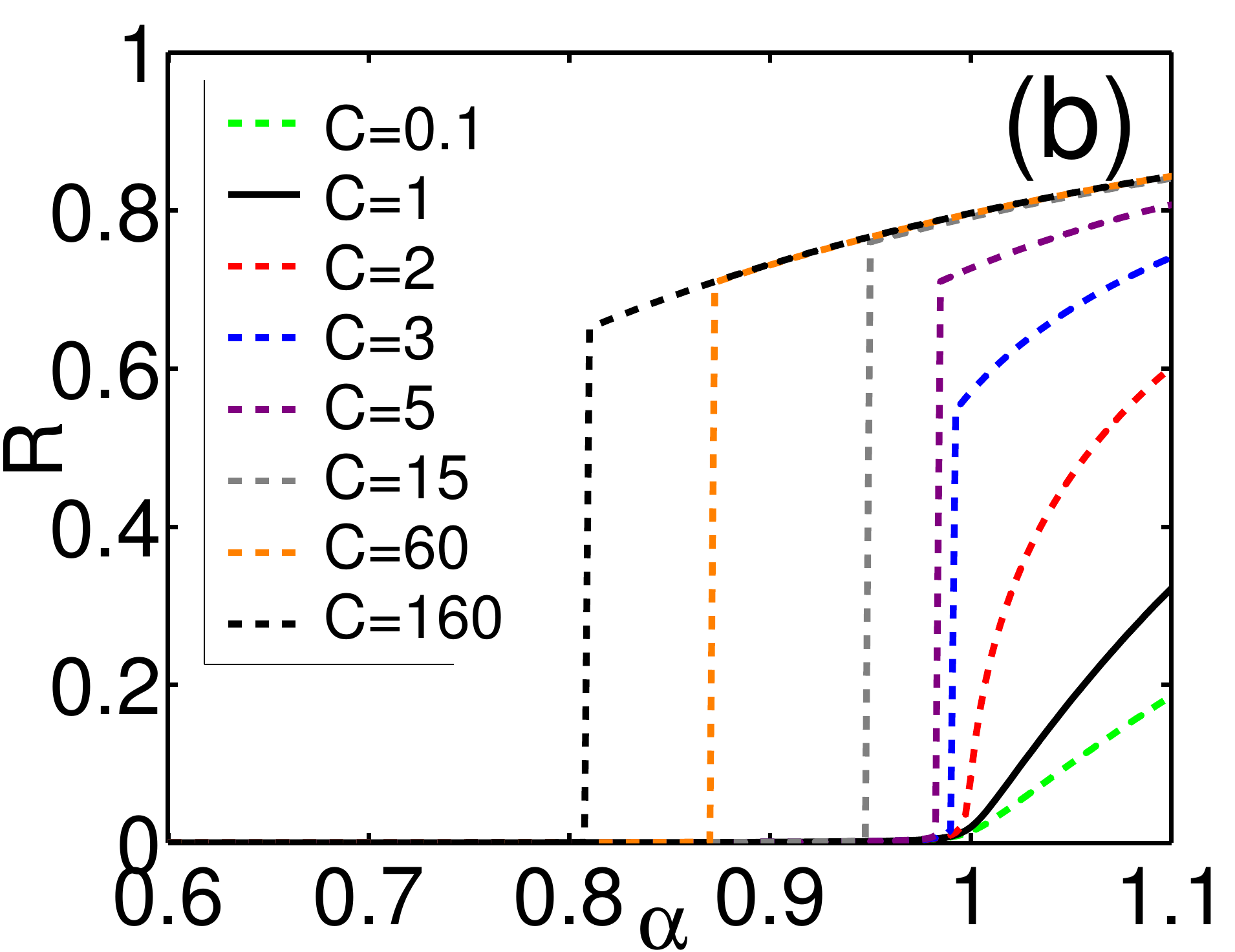}
\caption{(color online). Order parameter $R = 1-S_\infty$ plotted against $\alpha$ for (a) $\epsilon=0.005$
   and (b) $\epsilon=10^{-4}$. Each curve corresponds to a different level of cooperativity $C$. The
   dotted lines in panel (a) indicate the upper and lower limits $R^+$ and $R^-$ of the jumps at the first order
   transitions.
 }
\label{fig.2}
\end{figure}

(a) For $C <2$ and $\epsilon \to 0$ there can be epidemic outbreaks only when $\alpha >1$, corresponding
to the well known behavior of the single-disease SIR model. For $\alpha\approx1$, the order parameter grows linearly
with $\alpha$, $R\sim \alpha-1$, showing that the transition is continuous with order parameter exponent 1.
For $\epsilon >0$ the transition is rounded.

(b) When $C =2$ the transition is still continuous with threshold $\alpha^*=1$ (in the limit $\epsilon \to 0$),
but now the order parameter exponent is $1/2$.

(c) For $C > 2$ we observe first order transitions, when $1/2 < \alpha < 1$.
These transitions are sharp, even when $\epsilon>0$. On the other hand, when $\epsilon \to 0$ these transitions
occur for fixed $C$ at values $\alpha^*(C,\epsilon)$ that increase as $\epsilon\to 0$,
\be
    \lim_{\epsilon\to 0} \alpha^*(C,\epsilon) = 1             \label{alpha-star}
\ee
for any finite $C$. 

The behavior expressed in Eq.(\ref{alpha-star}) and illustrated in Fig. ~\ref{fig.2}(b) is an artifact of our
mean field approximation. Due to the latter, the cluster of infected
neighbors created by a sick individual is immediately dispersed in the entire population,
reducing thereby the chances for multiple infections. In any local model (i.e. on a regular lattice)
we would expect that this cluster stays localized for long time, so that even an infinitesimal
fraction of infective ``seeds" could lead to a large epidemic.

(d) Let us denote by $R^-(\alpha^*,C,\epsilon)$ and $R^+(\alpha^*,C,\epsilon)$ the lower and upper
values of the jumps at the first order transitions.  When $\alpha^*$ decreases to 1/2, they meet at
$R^\pm(1/2,C,\epsilon)$=0. When it increases, they both increase at first with $\alpha^*$. Later they meet, for
all finite $\epsilon>0$, at nontrivial values $\alpha_c(\epsilon)<1$
and $R_c(\epsilon)\in(0,1)$. At these points the transition is continuous, with the order
parameter exponent equal to 1/2.

(e) No epidemics are possible (for small $\epsilon$) when $\alpha < 1/2$, as also predicted
analytically by the theory discussed below.

(f) As long as $\alpha^* < \alpha < 1$, the values of $R$ are independent of $\epsilon$ within
numerical accuracy, but depend weakly on $C$. All values of $R$ are below the limit curve
\be
   R^+(\alpha) = \lim_{\epsilon\to 0}\lim_{C\to\infty} R(\alpha,C,\epsilon)
\ee
which scales as $R^+(\alpha) \sim \alpha-1/2$ for $\alpha\searrow 1/2$.

\begin{figure}[]
\includegraphics[scale=0.221]{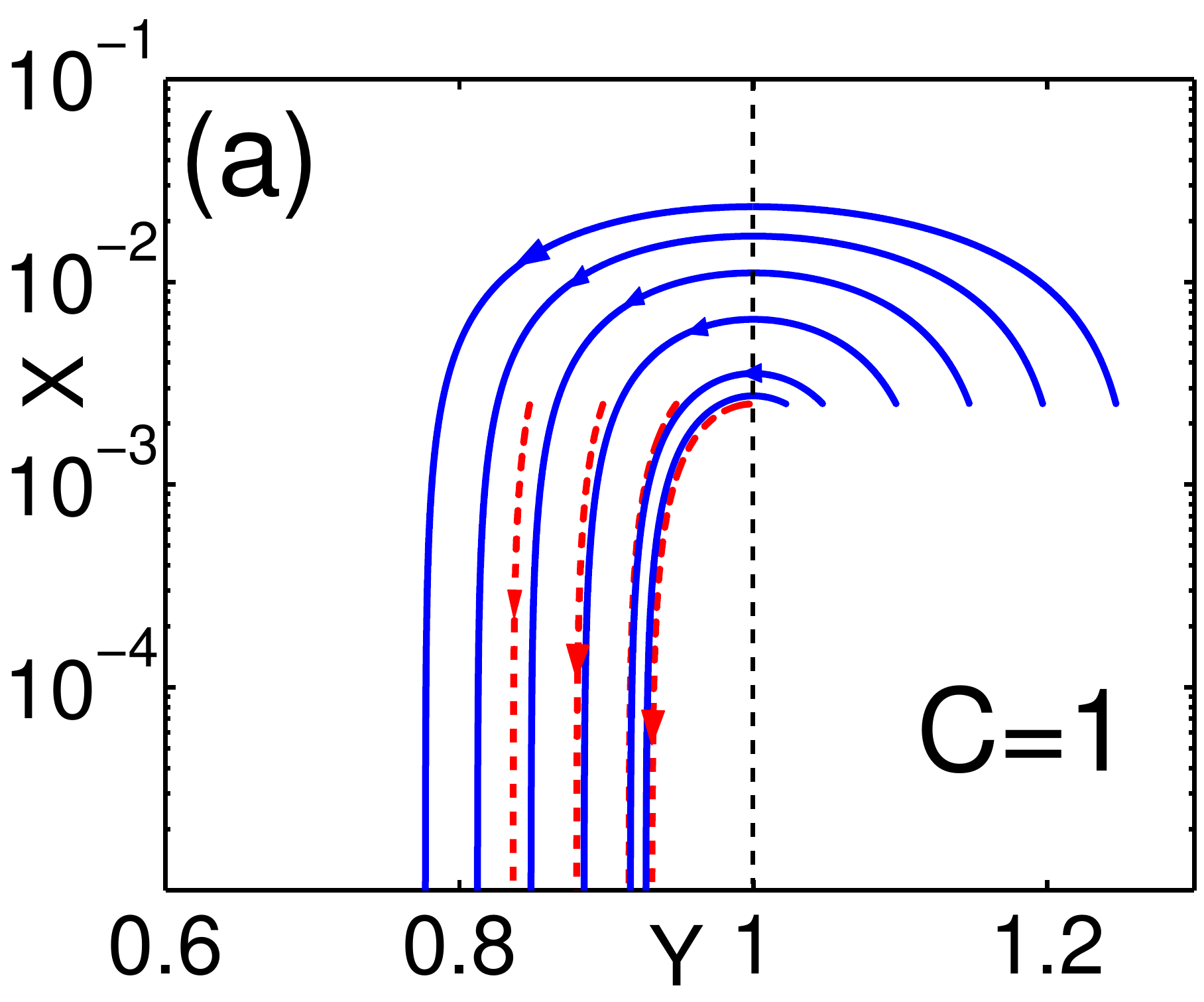}
\includegraphics[scale=0.221]{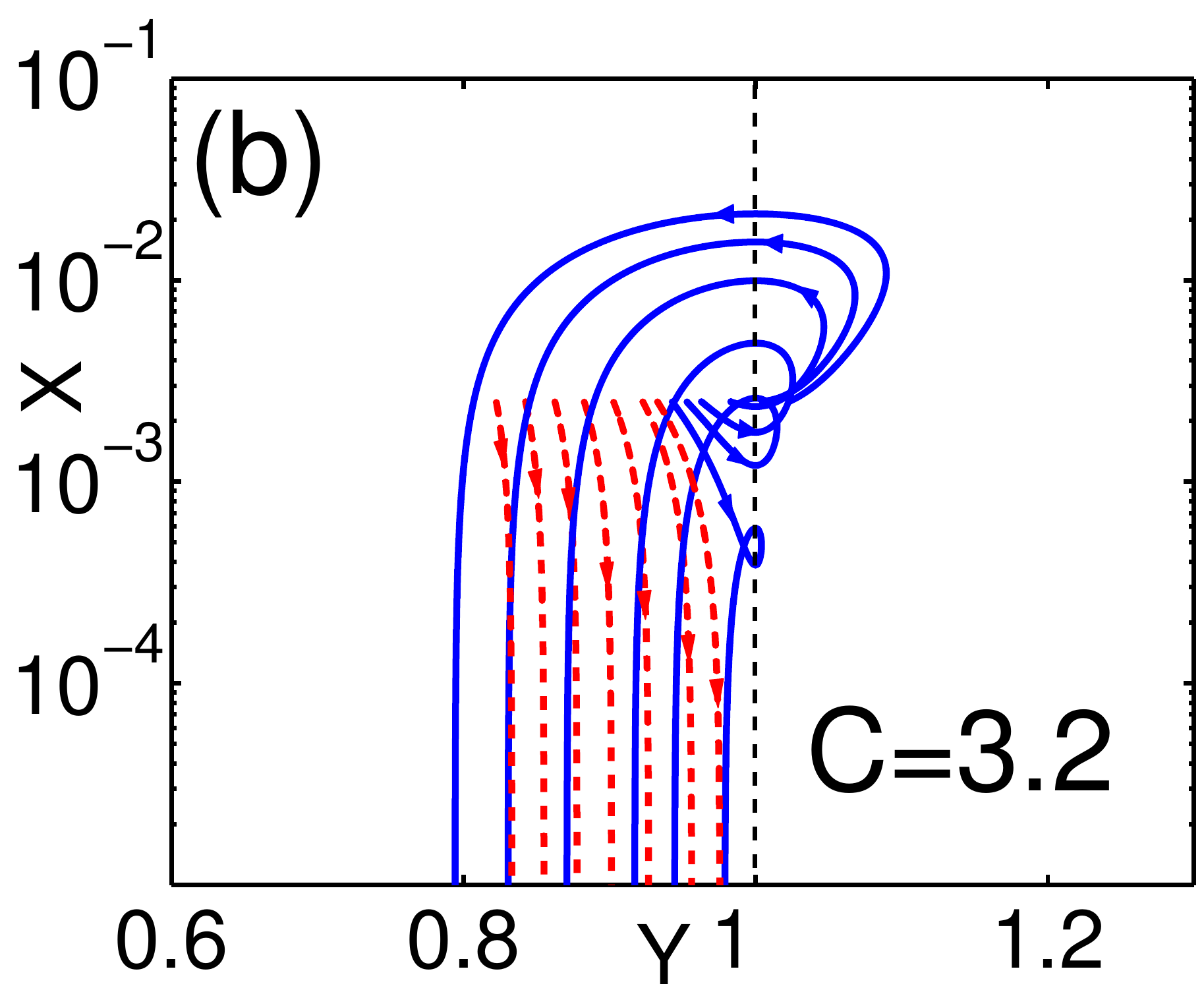}
\includegraphics[scale=0.221]{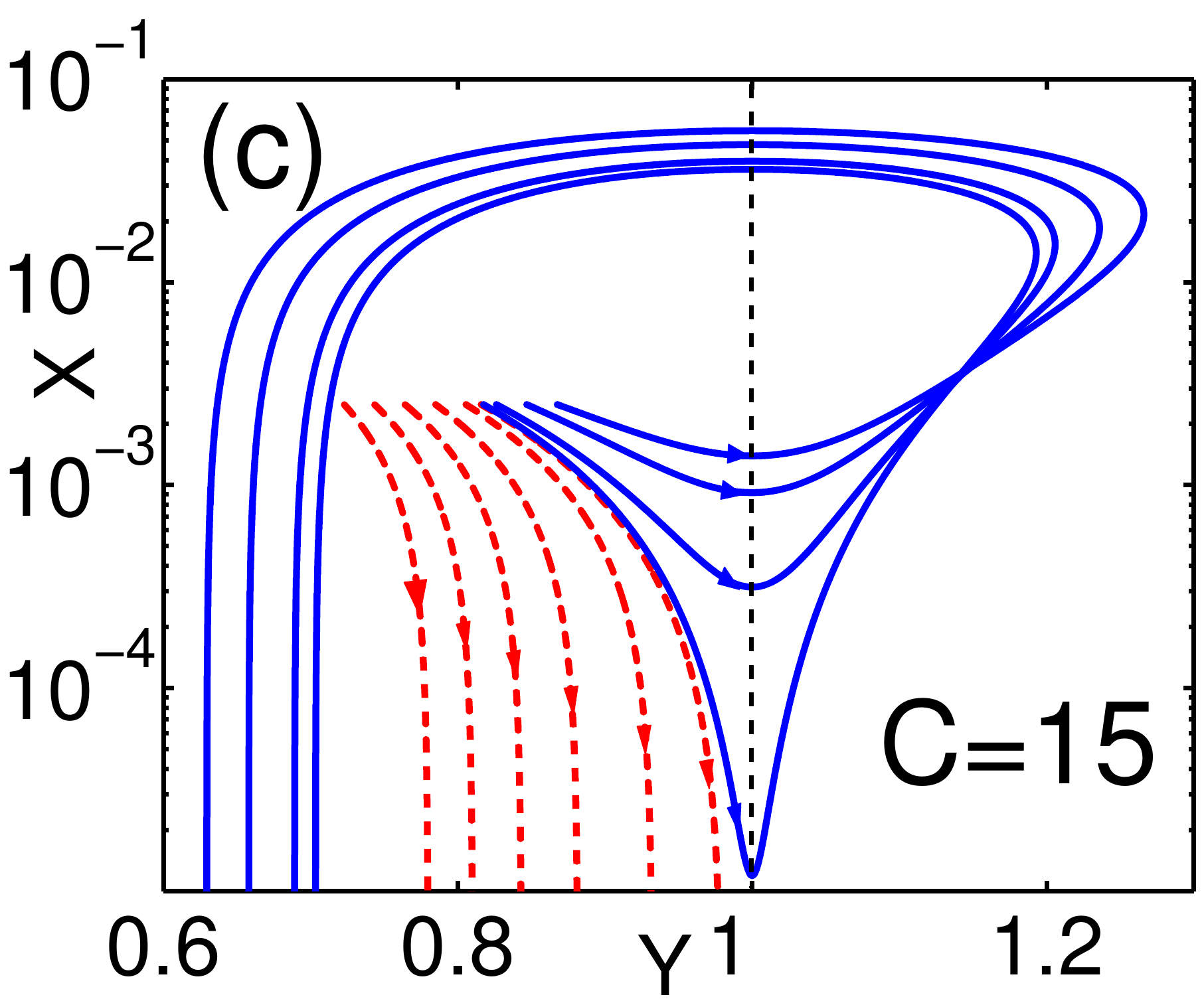}
\includegraphics[scale=0.221]{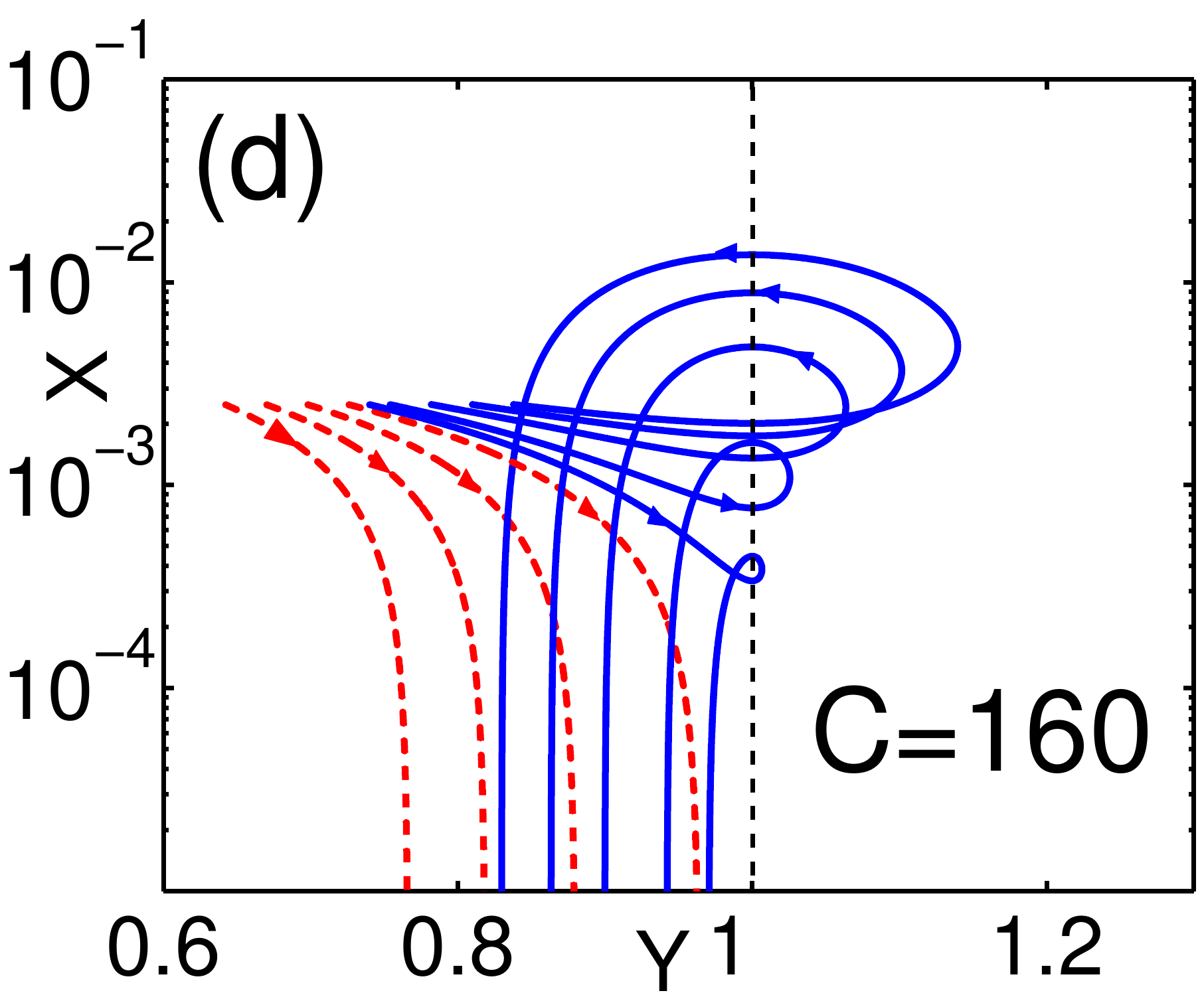}
\caption{(color online). Phase portraits of near-critical trajectories obtained by plotting
    $X(t)$ against $Y(t)$, for four different levels of cooperativities. Initial conditions 
    are the same as in Fig. 2(a).  Blue solid lines indicate cases with epidemic outbreaks 
    (supercritical), in contrast to red dashed lines (subcritical).
    The latter never cross the threshold $Y=1$ (black dash lines).
    Within each panel, different trajectories correspond to different infection rates $\alpha$.}
\label{traj}
\end{figure}

{\em Time dependence and theoretical explanations.} ---
In order to understand better the dynamics, we first define
$Y(t) = \alpha S(t) + \beta P(t)$,
whose time dependence is
\bea
    \dot{Y} &=& [(\beta-2\alpha)\alpha S -\beta^2 P]X        \label{Y1} \\
            &=& [2(\beta-\alpha)\alpha S - \beta Y]X          \label{Y2}
\eea

According to Eq.(\ref{ODE3}), $\dot{X} = (Y-1)X$. Therefore $X(t)$ can only grow when $Y(t)>1$.
But, due to Eq.~(\ref{Y1}), $Y$ can grow for small $\epsilon$ only iff $\beta
> 2\alpha$. This explains immediately why normal SIR threshold behavior is seen if
and only if $C <2$. Assume now that $ \alpha < 1$ and that $C$ is sufficiently large so that $(\beta-2\alpha)\alpha > 1$.
Then $Y$ will start to grow for sufficiently small $\epsilon$. If it grows to a value $1$, there
will be an outbreak. This might be prevented by two mechanisms:
Either $S$ decreases so fast and $Y$ increases so fast that the first factor on the r.h.s. of
Eq.~(\ref{Y2}) becomes zero, or $X$  -- the second factor in Eq.~(\ref{Y2}) -- vanishes. As
we shall see, these two alternatives give rise to first- and second-order phase transitions.

To proceed we use the exact inequality $\alpha S \leq Y$ in order to eliminate $S$ from Eq.~(\ref{Y2}),
and obtain for small times (as long as $Y<1$)
\be
    \frac{1-Y}{Y}\dot{Y} \leq -(\beta-2\alpha) \dot{X}.
\ee
This can be integrated to give an upper bound $X_+(Y)$ on $X$ that decreases monotonically with $Y$.
If $X_+(Y=1)<0$ , we know that there cannot be an outbreak. If $X_+(Y=1)>w$, where $w$ is a positive
constant independent of $\epsilon$, we must have a first order phase transition for sufficiently
small $\epsilon$ (where the inequality becomes practically tight), provided $\dot{Y}>0$ when $Y=1$.
Finally, if $X_+(Y=1)\geq 0$ but $\dot{Y}=0$ when $Y=1$, we have a second order transition.

These cases are illustrated in Fig.~\ref{traj}. In each panel of this figure, we show trajectories
of the flow by plotting $X(t)$ against $Y(t)$. Panel (a) shows a standard SIR
transition where the critical point corresponds to $\alpha=1$ and $Y$ decreases monotonically.
Panel (c) shows the generic case of strong cooperativity, where $Y$ increases beyond $Y=1$, provided that $X$ does
to go to zero before. If $Y$ passes through $Y=1$ it it continues to$Y \gg 1$ (even close to the transition
point), indicating a first order transition. Panels (b) and (d) show cases where $Y$ go only infinitesimally
beyond $Y=1$ at the transition point, corresponding to second order transitions. Panel (d) shows the
case of ultra-strong cooperativity, corresponding to the uppermost curve in Fig.~\ref{fig.2}(a), where
$\alpha^*= 1/2$. Panel (b), finally, corresponds to the special case of moderately weak cooperativity
where $R^-(\alpha^*,C,\epsilon) =R^+(\alpha^*,C,\epsilon)$, so that the jump hight in Fig.~\ref{fig.2}
just vanishes.

\begin{figure}[]
\includegraphics[scale=0.31]{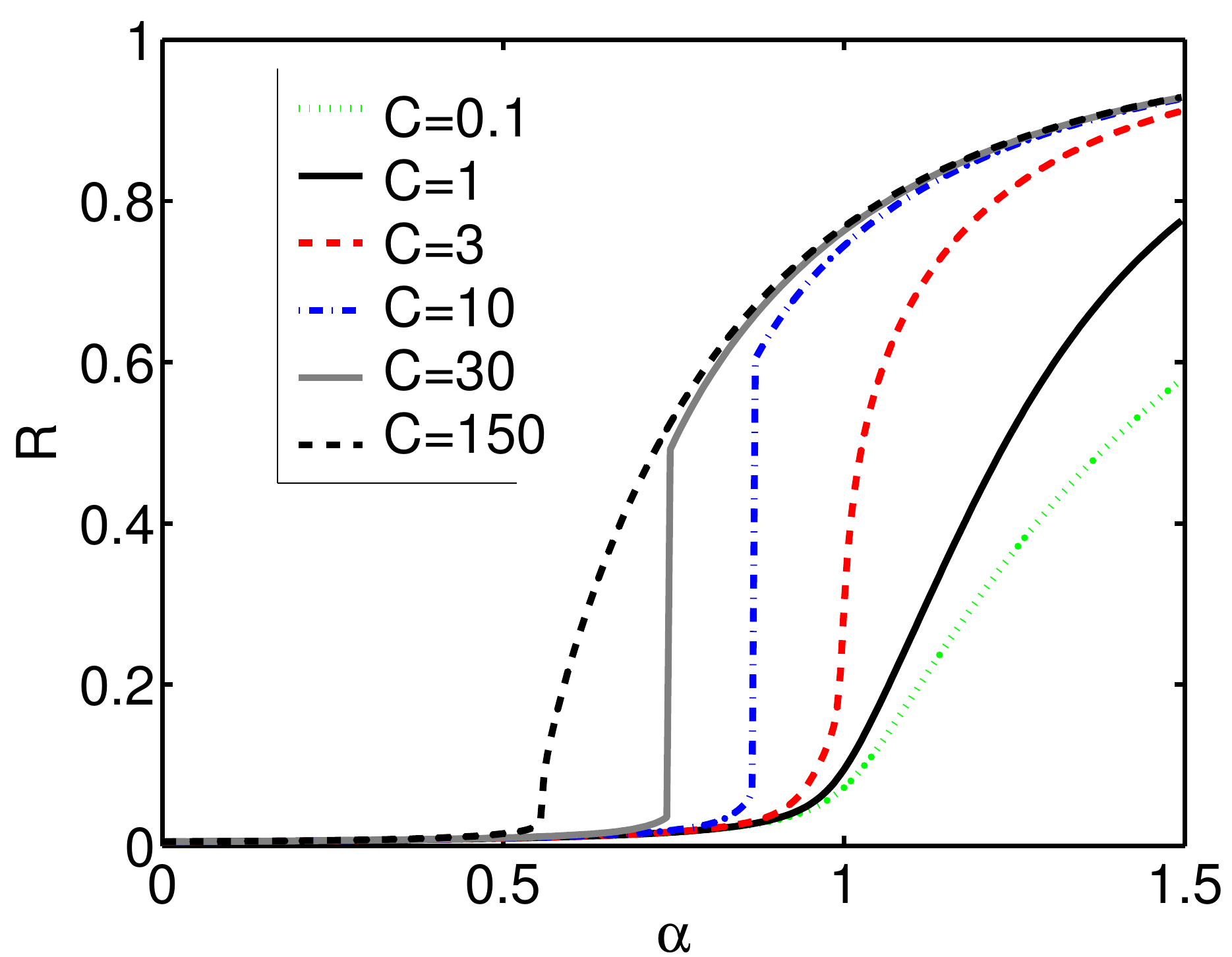}
\caption{(color online). Order parameter $R$ plot for a general case where both diseases have different infection
   rates, and where also the chances for getting infected depend on whether the target still has the other disease
   or has already recovered from it. More specifically we used initial conditions $[S]=0.995$, $[A]=0.002$, $[B]=0.003$,
   with infection rates $\alpha$ for $S\rightarrow A$; $\alpha'=0.9\alpha$ for $S\rightarrow B$; $\beta=C\alpha$
   for $A\to AB$; $\beta'=2C\alpha'$ for $B\to AB$; $\gamma=0.7\beta$ for $a\to aB$; and $\gamma'=0.8\beta'$ for
   $b\to Ab$.}                \label{fig:5}
\end{figure}

Up to now we have dealt with the special case with perfect symmetry between the two diseases, and
where the infection rate increase due to cooperativity is the same for targets that are still infected
and those which have already recovered from the other disease. In more general cases, where 
all parameters in Eq.(1) are different, we cannot give similarly detailed mathematical results, but we still
can make numerical simulations. We have found similar behaviors in all cases. 
One such case is shown in Fig.~\ref{fig:5}. There we still assume that all recovery rates
are equal, but all other symmetry restrictions are removed. We 
see the same type of phase transitions
as in Fig.~\ref{fig.2}. We thus conjecture that the behavior discussed above is indeed robust and prevails also
in more general cases.

\emph{Conclusions.} --- As we have shown, the cooperativity of coinfections can not only decrease the
thresholds for epidemic outbreaks, but it can also change the outbreak from continuous (``second order")
to discontinuous (``first order"). This may pose a much more serious problem in real situations. In second
order transitions the size of the epidemic grows gradually as conditions become more
favorable for an outbreak, and one has precursors
which may be used to initiate counter measures. In a first order transition
such precursors are absent, and the epidemic develops immediately its full size, once the threshold has
been overcome, leaving much less time to react. Intuitively, the discontinuity of the phase transitions 
results from the fact that the ``basic reproduction ratio" \cite{Anderson,Reluga} (which applies to infinitesimally
small initial epidemic seeds) is smaller than the reproduction ratio that applies when the fraction 
of infecteds is finite. 

Our results were only obtained in a very crude mean field treatment, and moreover our analytical results
dealt only with very special cases. But we checked numerically that they were robust in a wider setting,
and we conjecture that similar phenomena are seen when more sophisticated mathematical modeling is used,
such as spreading of the epidemics on spatial grids or methods similar to belief
propagation on (locally) loopless networks \cite{Newman,Goltsev,Woo}. Obviously much more work has to
be done, and the present letter should be seen only as a first small step towards mathematically
modeling more general and realistic situations.

In preliminary studies of a stochastic version \cite{tobepublished} we found no first order transitions 
on regular $d$-dimensional lattices in $d=2$ and $d=3$, if
infections are local (between nearest or next-nearest neighbors), but they do occur in $d=4$. They
also occur in $d=2$, if infection can happen with probability $P({\bf x})$ between nodes that are
a distance ${\bf x}$ apart, provided $P({\bf x})\sim |{\bf x}|^{-d-\sigma}$ for large ${\bf x}$
with small enough $\sigma$. As expected, we found first order transitions also in Erd{\"o}s-R{\'e}nyi (ER)
and small-world networks. In all these cases, we assumed that both diseases spread on the 
same set of links. If we had used two independent networks, spreading on ER networks would be 
identical to mean field. It is only the assumption that both diseases use the same network which 
makes spreading on ER networks different from mean field, and which allows epidemics in the first-order 
regime to spread already from infinitesimal seeds. 

Finally, we should point out that cooperative coinfections are not only important for
epidemiology in the narrow sense, but also for the spreading of computer malware, 
rumors, fashions, innovations, political opinions \cite{Lohmann} or social unrest \cite{Granovetter}.


\begin{thebibliography}{}

\bibitem{Hays} J. N. Hays, \emph{Epidemics and Pandemics: Their Impacts on Human History} 
   (ABC-CLIO, Santa Barbara, California, 2005).
\bibitem{Anderson} R. M. Anderson and R. M. May, \emph{Infectious diseases of humans: dynamics and control} 
    (Oxford University Press, Oxford New York, 1991).
\bibitem{Hethcote} H. W. Hethcote, SIAM Rev. {\bf 42}, 599 (2000)
\bibitem{KK} W. O. Kermack and A. G. McKendrick, Proceedings of the Royal Society A. {\bf 115}, 700 (1927).
\bibitem{Mollison} D. Mollison, J. Roy. Statist. Soc. {\bf B 39}, 283 (1977).
\bibitem{Grassberger} P. Grassberger, Math. Biosciences {\bf 63}, 157 (1983).
\bibitem{Newman} M. E. J. Newman, Phys. Rev. E {\bf 66}, 016128 (2002).
\bibitem{Dorogovtsev} S.N. Dorogovtsev, A.V. Goltsev, and J.F.F. Mendes, Rev. Mod. Phys. {\bf 80}, 1275 (2008).
\bibitem{Achlioptas} D. Achlioptas, R.M. D'Souza, and J. Spencer, Science {\bf 323}, 1453 (209).
\bibitem{Dodds} P. S. Dodds and D. J. Watts, Phys. Rev. Lett. {\bf92}, 218701 (2004); 
     J. Theor. Biology {\bf42}, 232 (2005).
\bibitem{Bizhani} G. Bizhani, M. Paczuski, and P. Grassberger, Phys. Rev. E {\bf  86}, 011128 (2012).
\bibitem{Goltsev} A. V. Goltsev, S. N. Dorogovtsev, and J. F. F. Mendes, Phys. Rev. E {\bf 73}, 056101 (2006).
\bibitem{Janssen} H.K. Janssen, M. M{\"u}ller, O. Stenull, Phys. Rev. E {\bf 70}, 026114 (2004).
\bibitem{Buldyrev} S.V. Buldyrev, R. Parshani, G. Paul, H.E. Stanley, and S. Havlin, Nature {\bf 464}, 1065, (2010).
\bibitem{Parshani} R. Parshani, S.V. Buldyrev, and S. Havlin, Phys. Rev. Lett. {\bf 105}, 048701 (2010).
\bibitem{Woo} S.-W. Son, G. Bizhani, C. Christensen, P. Grassberger, and M. Paczuski,
     Europhys. Lett. {\bf 97}, 16006 (2012).
\bibitem{Boettcher} S. Boettcher, V. Singh, and R.M. Ziff, Nature Commun. {\bf 3}, 787 (2012).
\bibitem{grass-levy-1d} P. Grassberger, J. Stat. Mech. P04004 (2013).
\bibitem{Reluga} T. C. Reluga, J. Medlock, and A. S. Perelson, J. Theor. Biol. {\bf 252}, 155 (2008).
\bibitem{Newman-a} M. E. J. Newman, Phys. Rev. Lett. {\bf 95}, 108701 (2005).
\bibitem{Funk10} S. Funk and V.A.A. Jansen, Phys. Rev. E {\bf 81}, 036118 (2010).
\bibitem{Marceau11} V. Marceau, P.-A. Noel, L. H{\'e}bert-Dufresne, A. Allard, and L.J. Dub{\'e},
    Phys. Rev. E {\bf 84}, 026105 (2011).
\bibitem{Miller} J.C. Miller, Phys. Rev. E{\bf 87}, 060801(R) (2013).
\bibitem{Singer09} M. Singer, {\it Introduction to Syndemics: A Critical Systems Approach to 
     Public and Community Health} (John Wiley \& Sons, 2009).
\bibitem{Brundage} J. F. Brundage and G. D. Shanks, Emerging Infectious Diseases, {\bf14}, 1193 (2008).
\bibitem{Oei} W. Oei and H. Nishiura, Comput. Math. Methods in Med., {\bf2012}, 124861 (2012).
\bibitem{Sulkowski} M. S. Sulkowski, J. Hepatology {\bf48}, 353 (2008).
\bibitem{Sharma} S. K. Sharma, A. Mohan, and T. Kadhiravan, Indian J. Med. Res. {\bf121}, 550 (2005).
\bibitem{Abu-Raddad} L. J. Abu-Raddad, P. Patnaik, and J.G. Kublin, Science {\bf 314}, 1603 (2006).
\bibitem{Marcheva} M. Martcheva and S.S. Pilyugin, SIAM J. Appl. Math. {\bf 66}, 843 (2006).
\bibitem{Pilyugin} S.S. Pilyugin, \url{http://mbi.osu.edu/2003/ws6materials/pilyugin.pdf} (2003).
\bibitem{Newman13}M. E. J. Newman and C. R. Ferrario, PLoS ONE 8(8): e71321 (2013).
\bibitem{tobepublished} L. Chen, F. Ghanbarnejad, W. Cai, and P. Grassberger, to be published.
\bibitem{Lohmann} S. Lohmann, World Politics {\bf47}, 42 (1994).
\bibitem{Granovetter} M. Granovetter, Am. J. Sociol. {\bf83}, 1420 (1978).

\end{thebibliography}
\end{document}